\documentstyle[12pt]{article}

\begin{document}
\title{\Large\bf Reconstructing azimuthal distributions in 
nucleus--nucleus collisions}
\author{Jean-Yves Ollitrault\footnotemark\\
Service de Physique Th\'eorique\footnotemark, CE-Saclay\\
91191 Gif-sur-Yvette, France\\ }
\maketitle
\footnotetext{$^*$ Member of CNRS.}
\footnotetext{$^\dagger$ Laboratoire de la Direction des Sciences de la 
Mati\`ere du Commissariat \`a l'Energie Atomique.}
\begin{abstract} 
Azimuthal distributions of particles produced in nucleus-nucleus collisions 
are measured with respect to an estimated reaction plane 
which, because of finite multiplicity fluctuations, differs 
in general from the true reaction plane. 
It follows that the measured distributions do not coincide with the true
ones. 
I propose a general method of reconstructing the Fourier coefficients
of the true azimuthal distributions from the measured ones. 
This analysis suggests that the Fourier coefficients are the best 
observables to characterize azimuthal anisotropies because, unlike
other observables such as the in-plane anisotropy ratio or the 
squeeze-out ratio, they can be reconstructed accurately. 
\end{abstract}
\newpage 
\setcounter{equation}{0} 
\parskip 6.0pt plus 2.0pt

A characteristic aspect of collective behavior in nucleus--nucleus
collisions is that 
the directions of the outgoing particles are correlated to the 
orientation of the impact parameter~\cite{bevalac}:  
azimuthal distributions measured from the reaction plane 
(which is the plane containing the impact parameter and the beam axis) 
are not uniform. 
At energies between 100~MeV and 2~GeV per nucleon, several experiments 
have measured azimuthal distributions of charged particles \cite{squeeze1}, 
identified protons and light nuclei \cite{fragment,fragment2}, 
charged \cite{kaos} and neutral \cite{pi0} pions, 
neutrons~\cite{neutrons,neutrons2} and $\Lambda$ baryons \cite{lambda}. 
Similar results have recently become available from the ultrarelativistic
heavy ion experiments performed at the Brookhaven AGS \cite{e877} and at 
the CERN SPS\cite{na49} where proton and charged pion azimuthal 
distribution have been measured. 
In all these analyses, the azimuthal angle is defined with respect 
to an estimated reaction plane which, because of finite multiplicity 
fluctuations, differs from the true reaction plane. 
It has been recently emphasized that a determination of nuclear equation 
of state from azimuthal anisotropies requires a good 
accuracy~\cite{fopi}. It is therefore important to correct 
the errors which are made in determining the reaction plane. 

We propose a systematic procedure to reconstruct the true 
azimuthal distribution from the measured ones. 
Normalized azimuthal distributions can be expressed as Fourier series:
\begin{equation}
\label{fourier}
{dN\over d\phi}={1\over 2\pi}\left( 1+2\sum_{n\ge 1} c_n\cos n\phi\right)
\end{equation}
where $c_n=\langle\cos n\phi\rangle$, 
the brackets denoting average values, and we assume that the 
azimuthal distributions are symmetric with respect to the reaction 
plane (i.e. even in $\phi$), which holds for spherical nuclei. 
The knowledge of all the Fourier coefficients 
$\langle\cos n\phi\rangle$ allows to reconstruct the full distribution 
using Eq.(\ref{fourier}). 

In an actual experiment, the reaction plane is not known exactly. 
It is reconstructed event by event from the reaction 
products. The reconstructed plane differs in general from the true
reaction plane by an error $\Delta\phi$, which varies from one event 
to the other. Thus, the measured azimuthal angle $\psi$ is related 
to the true azimuthal angle $\phi$ by $\psi=\phi-\Delta\phi$ (see Fig.1). 
Averaging over many events, assuming that $\phi$ and $\Delta\phi$ are 
statistically independent (this assumption will be discussed below), 
one obtains the following relation between the measured and true 
Fourier coefficients\cite{e877,demoulins,stara}:
\begin{equation}
\label{deconvolution}
\langle\cos n\psi \rangle  
=\langle \cos n\phi\rangle\langle \cos n\Delta\phi\rangle.
\end{equation}
From Eq.(\ref{deconvolution}), we can reconstruct the true distribution once 
the correction factor $\langle\cos n\Delta\phi\rangle$ is known. 

Before we calculate $\langle\cos n\Delta\phi\rangle$, let 
us briefly comment Eq.(\ref{deconvolution}). This equation shows that 
the measured anisotropies are always smaller than the true ones: 
they are smeared by the error $\Delta\phi$.  
More precisely, if the probability distribution of $\Delta\phi$ has 
a typical width $\delta$, i.e. if $\delta$ denotes the typical error 
made in determining the reaction plane, $\langle\cos n\Delta\phi\rangle$ 
will decrease with $n$ and become small for $n> 1/\delta$. 
This has two consequences. First, the higher order coefficients 
disappear in the measurement procedure. 
Indeed, all the distributions measured so far are well reproduced 
by keeping only the first two Fourier coefficients $n=1,2$ in 
Eq.(\ref{fourier}) \cite{squeeze1,kaos,pi0,neutrons,fopi}. 
However, higher order components might be sizeable in the true 
distributions. The second consequence is that observables involving 
higher order Fourier coefficients cannot be reconstructed accurately. 
This is unfortunately the case for widely used observables such as 
the in-plane anisotropy~\cite{neutrons,eos}
\begin{equation}
R_{\rm in-plane}={(dN/d\phi)_{\phi=0}\over(dN/d\phi)_{\phi=180^\circ}}
\end{equation}
or the squeeze-out ratio~\cite{squeeze1,kaos,pi0,neutrons2} 
\begin{equation}
R_{\rm squeeze}=
{(dN/d\phi)_{\phi=90^\circ}+(dN/d\phi)_{\phi=270^\circ}\over
(dN/d\phi)_{\phi=0^\circ}+(dN/d\phi)_{\phi=180^\circ}}
\end{equation}
which both involve an infinite number of Fourier coefficients 
(see Eq.(\ref{fourier})). 
Furthermore, Fourier coefficients, which are integrated quantities, 
are also easier to evaluate in theoretical models (especially in 
Monte-Carlo calculations) than observables using the value of the 
distribution at a specific point. 

We now recall how the orientation of the reaction plane is 
estimated from the reaction products:
in high energy collisions ($E/A>100$~MeV), the projectile (target) 
fragments are deflected away from the target (projectile).  
Therefore, the vector obtained by summing all the transverse momenta 
of the particles produced in the projectile (target) rapidity region 
is parallel (antiparallel) to the impact parameter. 
More generally, one constructs a vector {\bf Q}~\cite{odyniec}: 
\begin{equation}
\label{Q}
\mbox{\bf Q}=\sum_{k=1}^N{w_k\, {\bf u}_k}
\end{equation}
where the sum runs over all the detected particles in the event.  
${\bf u}_k$ is the unit vector parallel to the transverse momentum 
of the particle, and $w_k$ is a weight which may depend on the type of 
particle, its rapidity and transverse momentum. 
The choice of $w_k$ is to a large extent arbitrary. 
Danielewicz and Odyniec \cite{odyniec} choose $w_k=p_T$ 
for $y>0.3$, $w_k=-p_T$ if $y<-0.3$ and $w_k=0$ if $|y|<0.3$. 
Many alternative definitions have been used \cite{fai,beckmann}, 
some of which do not require particle identification. 
This method of determining the reaction plane is commonly referred to
as the transverse momentum method. 
In experiments which measure the energy deposited in a calorimeter, 
without counting individual particles, an equation similar to Eq.(\ref{Q})
must be used, with $w_k$ replaced by the energy deposited in each segment 
of the calorimeter \cite{voloshin}. 
We come back to this case at the end of this paper. 

For an ideal system with infinite multiplicity $N$, {\bf Q} lies in the true 
reaction plane, and azimuthal distributions can be measured from {\bf Q}. 
In an actual experiment, the multiplicity is finite, which has two effects. 
First, because of statistical fluctuations, there is a deviation 
$\Delta\phi$ between the true reaction plane and {\bf Q}. 
Another effect of the finite multiplicity is that when one measures the 
azimuthal angle of a particle with respect to {\bf Q}, there is a 
correlation if the particle is included in the sum in Eq.(\ref{Q}).
This can be avoided by constructing a new vector {\bf Q} 
obtained by summation over the  $N-1$ remaining particles \cite{odyniec}. 
Then, it is reasonable to assume, as we have done, that $\phi$ 
and $\Delta\phi$ are statistically independent. 
However, this is not quite true for small nuclei where the constraint 
of global momentum conservation creates important correlations. 
A method to subtract these correlations is described in \cite{constraint}. 

We now proceed to evaluate $\langle\cos n\Delta\phi\rangle$. 
We first show that the distribution of $\Delta\phi$ is a universal 
function of a single real parameter $\chi$, which measures the 
accuracy of the reaction plane determination, and scales with $N$ 
like $\sqrt{N}$. It is normalized in such a way that for large $\chi$, 
the standard angular deviation is 
$\langle\Delta\phi^2\rangle^{1/2}=1/(\chi\sqrt{2})$. 
Then we express $\langle\cos n\Delta\phi\rangle$ as a function of $\chi$. 
Finally, we explain how to extract $\chi$ from the data. 

We consider a large sample of events having 
the same magnitude of impact parameter. Experimentally, this can be done 
approximately by selecting events having the same multiplicity, 
or the same transverse energy, or the same energy in a zero degree calorimeter.
The number $N$ of particles entering the definition of the vector {\bf Q} 
in Eq.(\ref{Q}) is usually much larger than unity. 
Then the central limit theorem shows that, 
for given magnitude and orientation of the true impact parameter, 
the fluctuations of 
$\mbox{\bf Q}$ around its average value, 
$\langle{\bf Q}\rangle$, are gaussian. 
Note that both the magnitude and angle of the vector ${\bf Q}$ fluctuate 
(see Fig.2). We choose the direction of impact parameter as the $x$-axis. 
Then 
${\bf Q}=Q({\bf e}_x\cos\Delta\phi+{\bf e}_y\sin\Delta\phi)$
and $\langle{\bf Q}\rangle=\bar Q\mbox{\bf e}_x$, and the two dimensional 
distribution of ${\bf Q}$ takes the form
\begin{equation} 
\label{gaussian}
{dN\over Q dQ d\Delta\phi}={1\over\pi\sigma^2}
\exp\left(-{\left| {\bf Q}-\langle{\bf Q}\rangle\right|^2\over\sigma^2}\right)
={1\over\pi\sigma^2}
\exp\left(-{Q^2+\bar Q^2-2 Q\bar Q\cos\Delta\phi\over\sigma^2}
\right).
\end{equation}
We have assumed that the fluctuations are isotropic. 
This will be justified later. 
Note that $\bar Q$ scales like $N$ while $\sigma$ scales like $\sqrt{N}$. 

Eq.(\ref{gaussian}) can be easily integrated over $Q$~\cite{olli93} to 
yield the distribution of $\Delta\phi$:
\begin{equation}
\label{pphi}
{dN\over\Delta\phi}=
{1\over\pi}\exp(-\chi^2)\left\{ 1+z\sqrt{\pi}
\left[ 1+\mbox{\rm erf}(z)\right]\exp(z^2) \right\}.
\end{equation}
where $z=\chi\cos\,\Delta\phi$ and $\mbox{erf}(x)$ is the error function. 
This distribution depends on $\bar Q$ and $\sigma$ only through the 
dimensionless parameter $\chi\equiv\bar Q/\sigma$. 
The Fourier coefficients are most easily calculated by integrating 
Eq.(\ref{gaussian}) first over $\Delta\phi$ and then over $Q$ 
\cite{stara}:
\begin{equation}
\label{pn}
\langle\cos n\Delta\phi\rangle=
{\sqrt{\pi}\over 2}\chi e^{-\chi^2/2}\left[
I_{n-1\over 2}\left( {\chi^2\over 2}\right) +
I_{n+1\over 2}\left( {\chi^2\over 2}\right) \right]
\end{equation}
where $I_k$ is the modified Bessel function of order $k$.
The variations of the first coefficients with $\chi$ is displayed 
in Fig.3. As expected, $\cos n\Delta\phi$  decreases with increasing $n$,  
and becomes vanishingly small for $n\gg\chi$. 
Experiments report values of $\langle\cos\Delta\phi\rangle$ 
ranging from 0.35 \cite{demoulins} for light nuclei ($A\simeq 20$) 
to 0.94 \cite{kaos} for the heaviest ones ($A\simeq 200$), 
corresponding to values of $\chi$ between 0.4 and 2.2 respectively. 
Fig.3 shows that the corrections are important in this range. 

If $\chi\gg 1$, 
the distribution of $\Delta\phi$, Eq.(\ref{pphi}), becomes approximately 
gaussian 
\begin{equation}
\label{asymp2}
{dN\over d\Delta\phi}\simeq
{\chi\over\sqrt{\pi}}\exp\left(-\chi^2\Delta\phi^2\right)
\end{equation}
and the Fourier coefficients are given by 
\begin{equation}\label{asymp}
\langle\cos n\Delta\phi\rangle\simeq\exp(-n^2/4\chi^2).
\end{equation}

In the limit $\chi\ll 1$, the $n^{\rm th}$ Fourier coefficient
is of order $\chi^n$:
\begin{equation}
\label{asymp3}
\langle\cos n\Delta\phi\rangle\simeq{\sqrt{\pi}\over 2^n
\Gamma\left({n+1\over 2}\right)} \chi^n 
\end{equation}
where $\Gamma$ is the Euler function. 

We now turn to the determination of $\chi$. 
The most widely used method~\cite{kaos,pi0,neutrons,fopi}
to estimate the accuracy of the reaction plane determination 
is to divide each event randomly into two 
subevents containing half of the particles each, 
and to construct {\bf Q} for the two subevents \cite{odyniec}. 
One thus obtains two vectors $\mbox{\bf Q}_I$ and $\mbox{\bf Q}_{II}$. 
The distributions of $\mbox{\bf Q}_I$ and $\mbox{\bf Q}_{II}$ 
are given by an equation similar to Eq.(\ref{gaussian}). 
However, since each subevent contains only $N/2$ particles, 
the corresponding average value and fluctuations must be scaled:
$\bar Q_I=\bar Q_{II}=\bar Q/2$, $\sigma_I=\sigma_{II}=\sigma/\sqrt{2}$, 
and therefore $\chi_I=\chi_{II}=\chi/\sqrt{2}$. 
The distribution of the relative angle 
$\Delta\phi_R\equiv\left|\Delta\phi_I-\Delta\phi_{II}\right|$
can be calculated analytically 
(see the Appendix of \cite{olli93} and the note added in proof) 
\begin{eqnarray}
\label{corr}
{{\rm d}N\over{\rm d}\Delta\phi_R}={{\rm e}^{-\chi_I^2}\over 2}\left\{
{2\over\pi}(1+\chi_I^2)+
z\left[I_0(z)+\mbox{\bf L}_0(z)\right]
+\chi_I^2\left[I_1(z)+\mbox{\bf L}_1(z)\right]\right\}
\end{eqnarray}
where $z=\chi_I^2\cos\Delta\phi_R$ and $\mbox{\bf L}_0$ and $\mbox{\bf L}_1$ 
are modified Struve functions \cite{bateman}. 
This distribution is normalized to unity between $0$ and $\pi$. 

The value of $\chi$ can be obtained by fitting Eq.(\ref{corr})
to the measured distribution. 
However, it can be calculated more simply from the fraction 
of events for which $\Delta\phi_R>90^\circ$, which is calculated 
by integrating Eq.(\ref{corr}) over $\Delta\phi_R$: 
\begin{equation}\label{ratio}
{N(90^\circ<\Delta\phi_R<180^\circ)\over N(0^\circ<\Delta\phi_R<180^\circ)}
={\exp(-\chi_I^2)\over 2}={\exp(-\chi^2/2)\over 2}.
\end{equation}
Alternatively, one can obtain $\chi$ by measuring \cite{voloshin,olli93}
\begin{eqnarray}\label{cosn}
\langle\cos\Delta\phi_R\rangle &=&
\langle\cos\Delta\phi_I\rangle\langle\cos\Delta\phi_{II}\rangle\nonumber\\
&=& {\pi\over 8}\chi^2 e^{-\chi^2/2}
\left[ I_0\left(\chi^2/4\right)+I_1\left(\chi^2/4\right)\right]^2
\end{eqnarray}
where we have used Eq.(\ref{pn}) with $n=1$ and $\chi$ replaced by 
$\chi_I=\chi/\sqrt{2}$. 
The variation of $\langle\cos\Delta\phi_R\rangle$ with $\chi$ is 
displayed in Fig.4. 
Eq.(\ref{cosn}) is more reliable than 
Eq.(\ref{ratio}) if $\chi$ is large, for in this 
case the ratio in Eq.(\ref{ratio}) is very small and is therefore 
subject to relatively large statistical fluctuations. 
Other methods to measure $\chi$ are described in \cite{olli93,olli95}. 

We finally justify the hypothesis that was made in writing Eq.(\ref{gaussian}),
namely that the fluctuations have the same magnitude in both $x$ and $y$ 
directions. This is not a consequence of the central limit theorem, 
which only ensures that the two dimensional distribution of ${\bf Q}$ is 
gaussian. The most general gaussian compatible with the symmetry 
$\Delta\phi\rightarrow -\Delta\phi$ can be written as 
\begin{equation} 
\label{generalgaussian}
{dN\over Q dQ d\Delta\phi}={1\over\pi\sigma_x\sigma_y}
\exp\left[-{( Q\cos\Delta\phi-\bar Q)^2\over\sigma_x^2}
-{Q^2\sin^2\Delta\phi\over\sigma_y^2}\right].
\end{equation}
The quantities $\sigma_x$ and $\sigma_y$ characterize the fluctuations
along the $x$ and $y$ axis, which are {\it a priori\/} different 
(the circle in Fig.2 should then be replaced by an ellipse). 
$\bar Q$, $\sigma_x$ and $\sigma_y$ are related to the azimuthal 
distribution of particles in the following way. 
Assuming for simplicity that the multiplicity $N$ is the same for all 
events in the sample (it is at least approximately true since the impact 
parameter is fixed), we get from Eq.(\ref{Q})
\begin{equation}
\label{barQ}
\bar Q= \langle \mbox{\bf Q}\cdot\mbox{\bf e}_x \rangle
=N \langle w\cos\phi\rangle 
\end{equation}
where $\phi$ is the {\sl true} azimuthal angle of particles, and 
the last average involves all the detected particles of all events. 
Similarly, the fluctuations in the $x$ and $y$ directions are given by 
\begin{eqnarray}
\label{sigmaxy}
\sigma_x^2&=&
2\left[\langle (\mbox{\bf Q}\cdot\mbox{\bf e}_x)^2 \rangle-\bar Q^2\right]
=2N\left[ \langle w^2\cos^2\phi\rangle 
-\langle w\cos\phi\rangle ^2\right]\cr
\sigma_y^2&=&2\langle (\mbox{\bf Q}\cdot\mbox{\bf e}_y)^2 \rangle
=2N\langle w^2\sin^2\phi\rangle .
\end{eqnarray}
We define the average fluctuation 
$\sigma$ by 
\begin{equation}
\label{sigma}
\sigma^2 ={1\over 2}(\sigma_x^2+\sigma_y^2)=\langle Q^2\rangle -\bar Q^2
= N\left[ \langle w^2 \rangle -\langle w\cos\phi\rangle ^2\right]
\end{equation}
and the anisotropy of the fluctuations by
\begin{equation}
\label{sigmaa}
{1\over 2}(\sigma_x^2-\sigma_y^2)=N\left[\langle w^2\cos 2\phi\rangle
-\langle w\cos\phi\rangle^2\right].
\end{equation}
Three cases must be distinguished, depending on the relative magnitudes
of $\bar Q$, $\sigma_x$ and $\sigma_y$. 
\begin{itemize}
\item (A) 
Azimuthal anisotropies are small, i.e. the Fourier coefficients of the 
azimuthal distribution are much smaller than unity. Then 
$\langle w^2\cos 2\phi\rangle$ and $\langle w\cos\phi\rangle^2$ 
are both small compared to $\langle w^2\rangle$. We deduce from 
Eqs.(\ref{sigma}) and (\ref{sigmaa}) that 
$\sigma_x\simeq\sigma_y\simeq\sigma$:  our assumption is justified in 
this case. 
\item (B) There are situations where $\langle w\cos\phi\rangle^2$ 
cannot be neglected compared to $\langle w^2\rangle$. 
This is the situation when the flow is strong. Using Eqs.(\ref{barQ}) 
and (\ref{sigmaxy}) and the fact that $N\gg 1$, we see that in this case
$\bar Q\gg\sigma_x,\sigma_y$: fluctuations are small and 
$\Delta\phi\simeq\mbox{\bf Q}\cdot\mbox{\bf e}_y/\bar Q\ll 1$ (see Fig.2). 
Therefore, 
the distribution of $\Delta\phi$ involves only $\sigma_y$, not $\sigma_x$.
Although $\sigma_x$ and $\sigma_y$ may differ in this case, one can 
replace $\sigma_x$ by $\sigma_y$ without altering the distribution of 
$\Delta\phi$. Note that in this case, $\chi=\bar Q/\sigma \gg 1$, 
hence the distribution of $\Delta\phi$ reduces to its asymptotic form,
Eq.(\ref{asymp2}). 
\item (C) Finally, there is the case when 
$\langle w^2\cos 2\phi\rangle$ is of order $\langle w^2\rangle$ while 
$\langle w\cos\phi\rangle^2$ is much smaller. 
In this case, there is a strong anisotropy in the second Fourier 
component, which should be used to determine the reaction plane. 
Instead of ${\bf Q}$, one should construct 
the vector $\mbox{\bf Q}_2$ defined as 
\begin{equation}
\label{Q2}
\mbox{\bf Q}_2=\sum_{k=1}^N{w^\prime_k\left(\mbox{\bf e}_x\cos 2\phi_k+
\mbox{\bf e}_y\sin 2\phi_k\right)}
\end{equation}
with the same notations as in Eq.(\ref{Q}), and $w^\prime_k$ is an 
appropriate weight. 
The azimuthal angle of the reaction plane is estimated as 
half the azimuthal angle of $\mbox{\bf Q}_2$, 
hence it is defined modulo $\pi$, i.e. one cannot distinguish 
$\phi$ and $\phi+\pi$.  
This method is equivalent to the diagonalisation of the 
transverse sphericity tensor, which has been claimed to be more efficient 
than the transverse momentum method at intermediate energies 
($E/A < 100$~MeV, see \cite{2dsphericity} and in particular Eq.(12) 
of \cite{tsang1}) and at ultrarelativistic energies 
where the flow angle is very small due to increasing 
nuclear transparency \cite{olli93,olli92}. 
Then, the procedure described in this paper can be applied to reconstruct 
the azimuthal distributions, replacing ${\bf Q}$ by ${\bf Q_2}$. 
However, since the azimuthal angle is defined modulo $\pi$, 
only the even Fourier components can be reconstructed. 
\end{itemize}

Before we conclude, let us briefly comment on corrections which have 
been applied, in previous works, to the measured azimuthal distributions. 
The method proposed by Tsang {\it et al.\/} \cite{tsang1,tsang2} is based on 
an ansatz for the distributions of $\phi$ and $\Delta\phi$, which are 
assumed to be proportional to $\exp(-\omega^2\sin^2\phi)$, where 
$\omega$ is a fitted parameter. 
Our analysis is more general in the sense that it does not make any 
a priori hypothesis on the shape of azimuthal distributions.
The analysis of Demoulins {\it et al.\/} \cite{demoulins} is similar 
to ours, although limited to $n=1$ and $n=2$. The main difference is 
that they calculate $\langle\cos\Delta\phi\rangle$ and 
$\langle\cos 2\Delta\phi\rangle$ independently from measured quantities. 
The analysis done by the E877 collaboration \cite{e877} is similar, in the 
sense that the corrections $\langle\cos n\Delta\phi\rangle$ are 
extracted directly, for each value of $n$, from measured correlations 
between subevents. 
On the contrary, our method allows to calculate simply all the Fourier 
coefficients $\langle\cos n\Delta\phi\rangle$ from Eq.(\ref{cosn}) or 
Fig.3, as soon as we know the first one, $\langle\cos\Delta\phi\rangle$. 

In conclusion, we have described a simple procedure to 
reconstruct the true azimuthal distributions from the measured ones 
by means of analytical formulae. Let us summarize this procedure: 
given a large sample of events in a restricted centrality interval, 
one measures the distribution of the relative angle between subevents 
$\Delta\phi_R$, from which one extracts, using Eq.(\ref{ratio}) or 
Eq.(\ref{cosn}), the crucial parameter $\chi$ which measures 
the accuracy of the reaction plane determination. 
Then one uses Eqs.(\ref{deconvolution}) and (\ref{pn}) to reconstruct
the Fourier coefficients of the true azimuthal distribution from the 
measured ones. 

A first assumption, on which our calculation relies crucially, is that 
the fluctuations of the momentum transfer ${\bf Q}$ are gaussian. 
This is reasonable only if the multiplicity is large enough. 
The method should not be applied if, for instance, a single big fragment of 
the projectile gives the largest contribution to ${\bf Q}$. In order to test 
the validity of the gaussian hypothesis, one may check that Eq.(\ref{corr}) 
provides a good fit to the measured distribution of the angle $\Delta\phi_R$. 
Alternative tests are proposed in \cite{olli95}. 
Our second assumption was that the subevents used in determining the 
value of $\chi$ are equivalent. This holds strictly only if the 
subevents are defined by a random selection of $N/2$ particles. 
The subevents are not equivalent if they correspond to different 
pseudorapidity windows as in \cite{e877}. However, Eq.(\ref{cosn}) 
can still be used to calculate the correction for arbitrary $n$, 
once the correction for $n=1$ is known. 

We have introduced a parameter $\chi$ to measure the accuracy of the 
reaction plane determination. 
Unlike other quantities frequently used in the literature such as 
$\langle\cos\Delta\phi\rangle$~\cite{fragment2,demoulins}, 
or $\langle\Delta\phi^2\rangle^{1/2}$~\cite{kaos,pi0,neutrons,lambda}, 
$\chi$ scales simply with the multiplicity like $\sqrt{N}$. 
It has a simple physical interpretation, being the ratio of the 
average value of the flow vector ${\bf Q}$ to 
the typical statistical fluctuation $\sigma$. And furthermore, it 
can be directly deduced from measured quantities by simple 
expressions such as Eq.(\ref{ratio}). 

Our analysis suggests that the Fourier coefficients of the azimuthal 
distribution, which can be reconstructed accurately and are also easy 
to estimate in theoretical models, are the best 
observables to characterize azimuthal anisotropies. 
Note that we have chosen to measure azimuthal angles around the beam axis. 
It was argued in Ref.\cite{squeeze2} that 
azimuthal distributions should rather be measured around the flow axis, 
determined by a sphericity tensor analysis \cite{sphericity}. 
But the fluctuations of the sphericity tensor, which is a $3\times 3$ matrix, 
are much more complex \cite{gyulassy} than those of {\bf Q}, and 
cannot be described in terms of a single parameter $\chi$. 
There exists no simple procedure to subtract statistical errors 
in this case. 

What accuracy can be attained with our method? 
If the azimuthal angles of ${\cal N}$ particles 
are measured (summing over all events), the corresponding statistical 
error on the measured Fourier coefficient is $1/\sqrt{2{\cal N}}$. 
Systematic errors can be removed, at least partly, using a mixed event 
technique. The error on the true Fourier coefficient will be larger 
by a factor $1/\langle\cos n\Delta\phi\rangle$, according to 
Eq.(\ref{deconvolution}). 
However, with high enough statistics, accurate measurements are possible 
even if the reaction plane is poorly known, as is 
the case with light projectiles \cite{demoulins}. 
In experiments using heavy ions, it would be interesting to try to 
measure higher order Fourier coefficients with $n\ge 3$ which, although
probably small, could be measured accurately. 

\section*{Acknowledgements} 
I thank B\'eatrice de Schauenburg and Philippe Crochet for discussions, 
and Jean-Paul Blaizot for useful comments on the manuscript,

\section*{Figure captions}

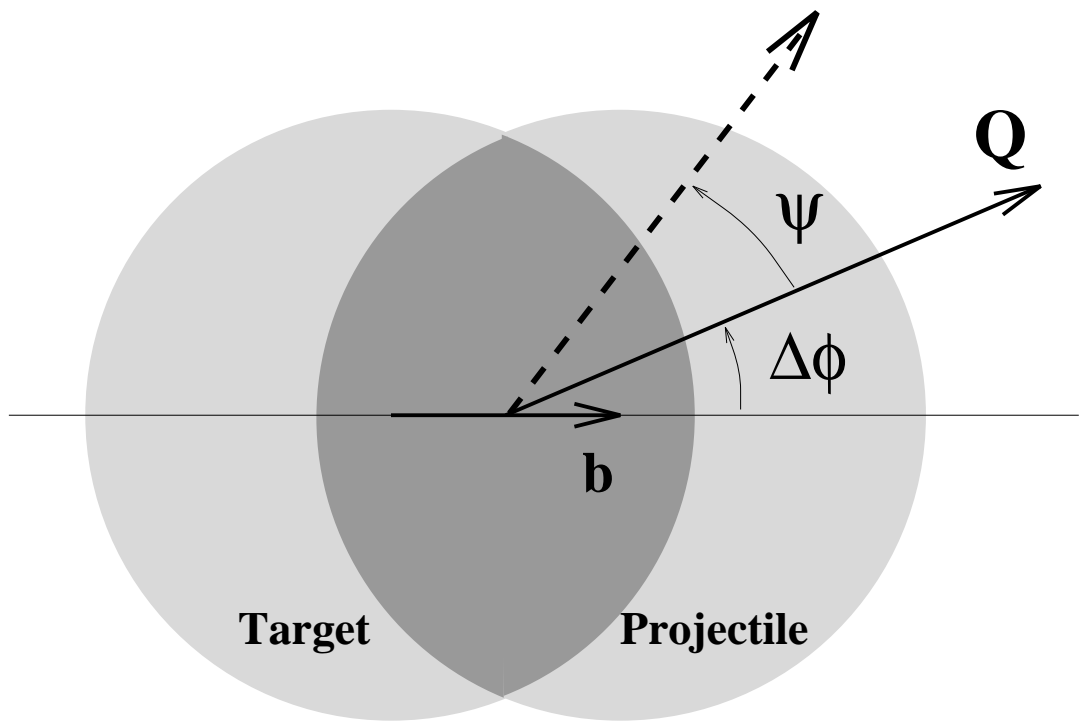
\noindent{\bf Fig. 1:} Schematic picture of a semi-central nucleus-nucleus
collision viewed in the transverse plane (the beam axis is orthogonal to the 
figure). 
{\bf b} is the impact parameter oriented from the target to the projectile 
and {\bf Q} is the vector defined by Eq.(\ref{Q}). 
A particle is emitted along the dashed arrow. 
Its azimuthal angle measured with respect to {\bf Q} is $\psi$, while 
the ``true'' azimuthal angle is $\psi+\Delta\phi$. 

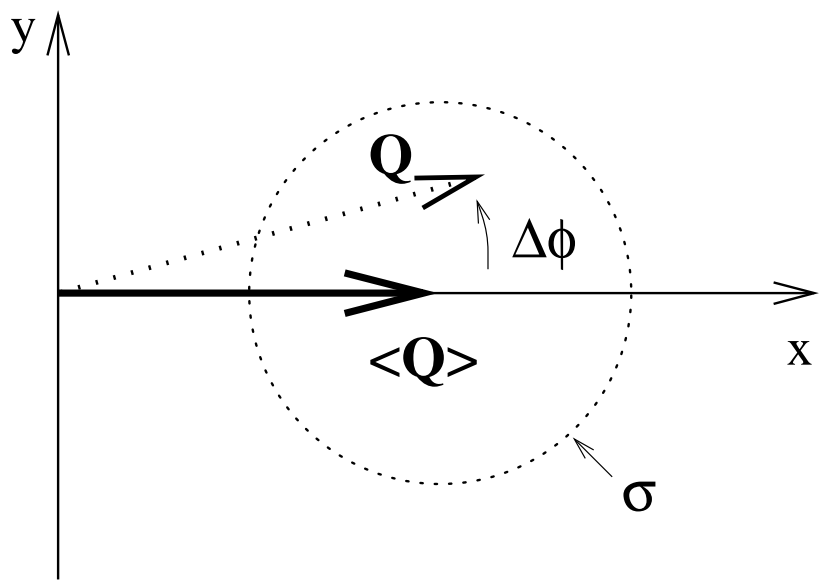
\noindent{\bf Fig. 2:} Schematic picture of the distribution of ${\bf Q}$, 
given by Eq.(\ref{gaussian}). 
The solid thick arrow indicates the average value 
$\langle {\bf Q}\rangle=\bar Q {\bf e}_x$, 
which lies along the direction of the true impact parameter. 
${\bf Q}$ fluctuates around this average value with a standard deviation 
$\sigma$, so that a typical value of {\bf Q} (dotted arrow) 
lies within the dotted circle with radius $\sigma$. 
It is obvious from this figure that the typical magnitude of $\Delta\phi$ is 
$\sigma/\bar Q=1/\chi$. 

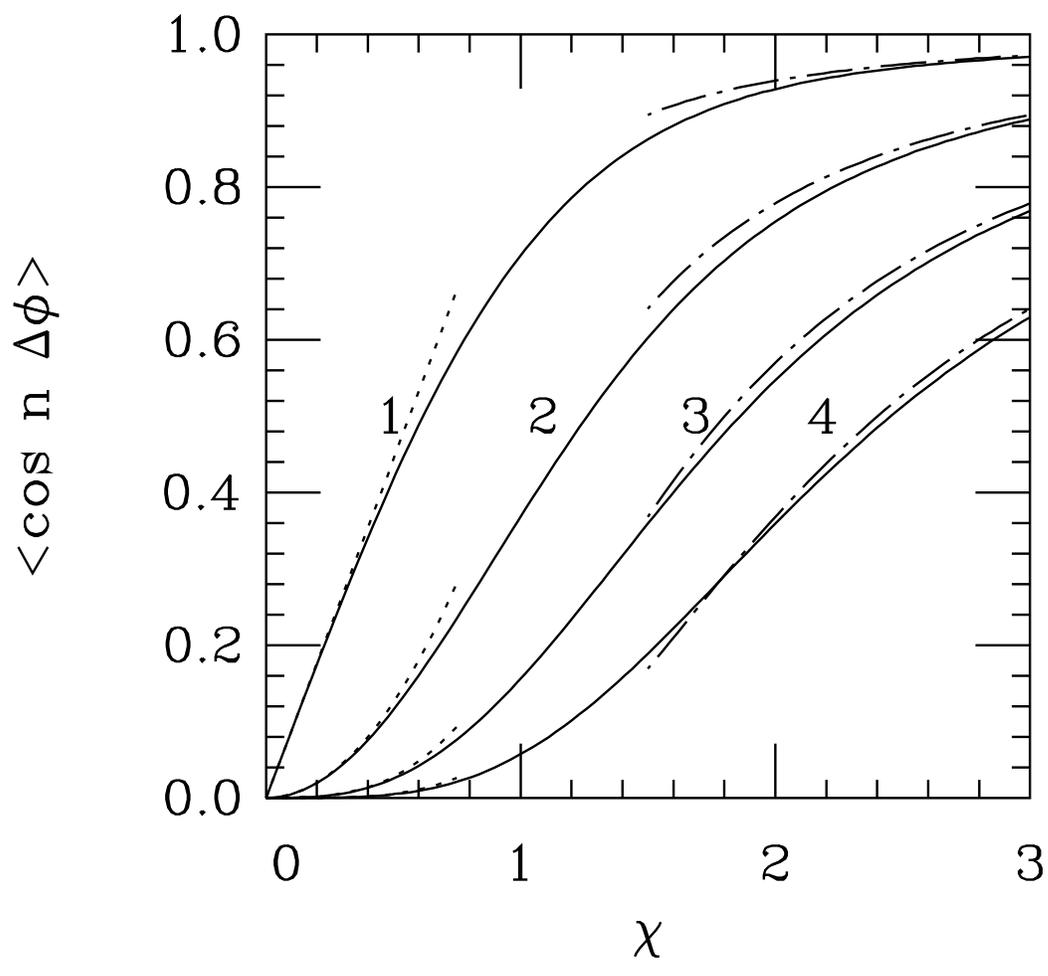
\noindent{\bf Fig. 3:} Solid lines: variation of 
$\langle\cos n\Delta\phi\rangle$ with the parameter $\chi$, calculated 
from Eq.(\ref{pn}). The curves are labeled by the value of $n$. 
The dotted curves and dash-dotted curves are the asymptotic forms 
given respectively by Eqs.(\ref{asymp3}) and (\ref{asymp}).  

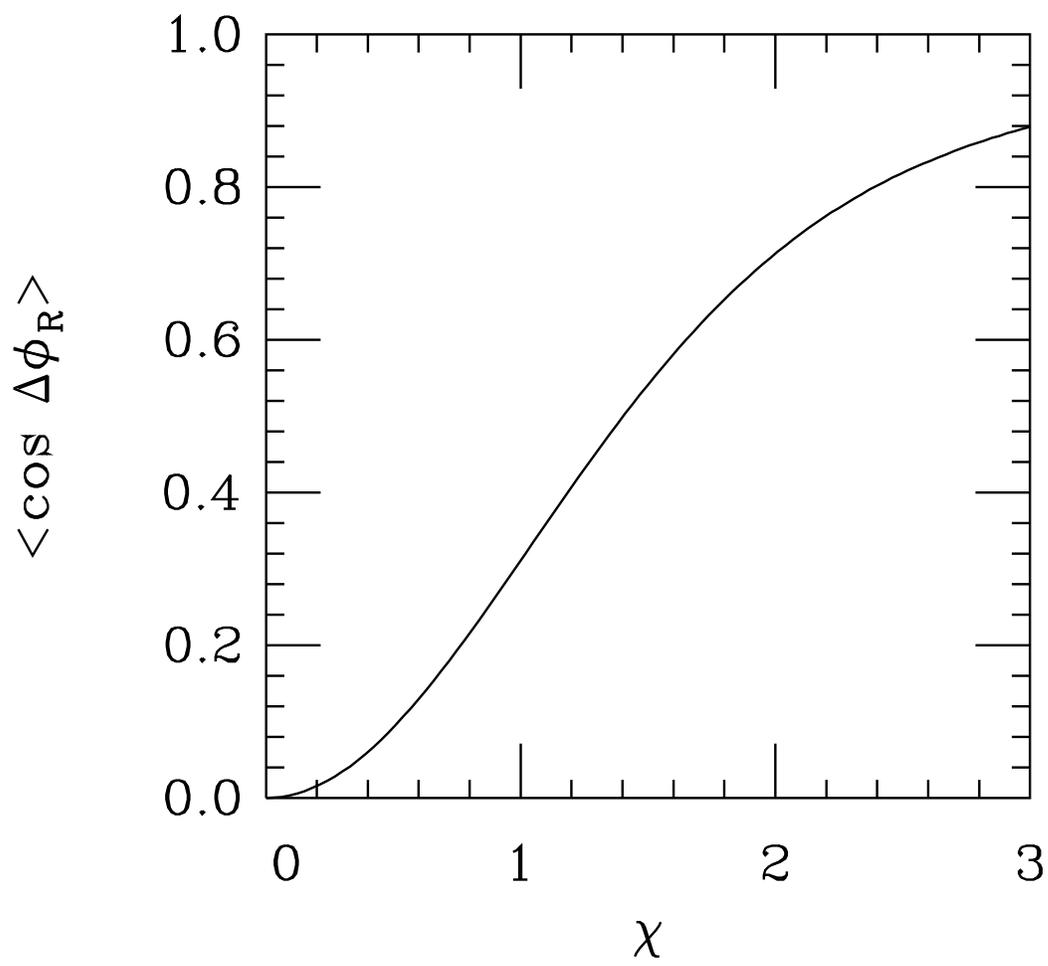
\noindent{\bf Fig. 4:} 
Variation of $\langle\cos\Delta\phi_R\rangle$ with $\chi$, 
given by Eq.(\ref{cosn}).

\end{document}